\documentclass[aps,twocolumn,
showpacs,groupedaddress,floatfix]{revtex4}
\usepackage[dvips]{epsfig}

\begin{document}

\title{Electronic structure of the molecule based magnet
Cu PM(NO$_3$)$_2$  (H$_2$O)$_2$}

\author{K. Doll}
\affiliation{Institut f\"ur Mathematische Physik, TU Braunschweig, 38106
Braunschweig, Germany\\
Max Planck Institute for Solid State Research, Heisenbergstr. 1, 70569
Stuttgart, Germany}

\author{A. U. B. Wolter}
\affiliation{Institut f\"ur Physik der Kondensierten Materie, TU Braunschweig,
38106 Braunschweig, Germany\\
Hahn-Meitner-Institut at BESSY, Albert-Einstein-Stra{\ss}e 15, 
12489 Berlin, Germany}

\author{H.-H. Klauss}
\affiliation{Institut f\"ur Physik der Kondensierten Materie, TU Braunschweig,
38106 Braunschweig, Germany}

\date{\today}

\begin{abstract}
We present density functional calculations on the molecule based 
$S=\frac{1}{2}$  antiferromagnetic chain compound
Cu PM(NO$_3$)$_2$  (H$_2$O)$_2$; PM = pyrimidine. The
properties of the ferro- and antiferromagnetic state are investigated
at the level of the local density approximation and with the hybrid functional
B3LYP. Spin density maps illustrate the exchange path via the pyrimidine
molecule which mediates
the magnetism in the one-dimensional chain. The computed exchange
coupling is antiferromagnetic and in reasonable agreement with
the experiment. It is suggested that the antiferromagnetic coupling
is due to the possibility of stronger delocalization of the
charges on the nitrogen atoms, compared to the ferromagnetic case.
In addition, computed isotropic and
anisotropic hyperfine interaction parameters are compared with recent
NMR experiments.
\end{abstract}


% 75.30.Et Exchange and superexchange interactions
% 75.50.Ee Antiferromagnetics
% 75.50.Xx Molecular magnets

\maketitle

\section{Introduction}

Molecule based magnetic materials have been the subject of intense research,
with the target of designing new magnetic materials\cite{KahnBuch}.
For example, novel quantum phenomena
such as quantum tunneling of magnetization 
open up possible future applications in
quantum computing and data storage
\cite{Friedman1996,Thomas1996,Barbara1999,Gatteschi2003}.
Other interesting research directions are spin state transitions
\cite{Guetlich}, or tuning of the magnetic coupling and the
design of ferromagnets (see, e.g. \cite{Verdaguer2001,pilawa1999}). 
This has led to an interdisciplinary effort in physics and chemistry,
and by experimentalists and theoreticians.

Supramolecular complexes of transition metals with organic ligands 
can also be used to synthesize low-dimensional magnets.
The organic ligands constitute
magnetic superexchange pathways with a strength 
of the order of 1 to 100 Kelvin. Since pyrimidine and similar
heterocycles (pyrazine, pyridine) are often found as magnetic
exchange mediating molecules in metal-organic magnets (e.g.
\cite{Kreitlow,Clerac2002,Hammar1999}), it is important to
study in detail the electronic structure and magnetic exchange mechanism.

\begin{figure}[ht]
\begin{center}
\includegraphics[width=9cm]{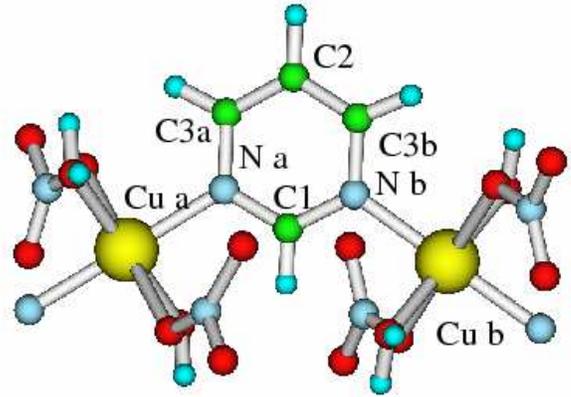}
\caption{\label{cupmbild} A cutout of the one dimensional chain of 
Cu PM(NO$_3$)$_2$  (H$_2$O)$_2$.}
\end{center}
\end{figure}

Cu PM(NO$_3$)$_2$  (H$_2$O)$_2$ is
a molecule based magnet which can be considered as 
a one dimensional spin chain (see figure \ref{cupmbild}). 
This complex was synthesized a few years ago \cite{Ishida1997}. It
has been studied by 
magnetic susceptibility, specific heat and electron spin 
resonance measurements \cite{Feyerherm2000}. More recently,
high field magnetization studies \cite{anja2003prb} and $^{13}$C NMR
measurements were performed \cite{anja2003poly,anja2005prl}.
It can be theoretically 
described as a $S=\frac{1}{2}$ antiferromagnetic
Heisenberg chain with an exchange coupling of 
$J$=36 K \cite{Feyerherm2000,Yasui2001,anja2003prb}, and 
with an additional Dzyaloshinskii-Moriya interaction and
a staggered g-tensor \cite{Oshikawa1997,anja2005prl}. This model had also
been used for other one-dimensional spin chain systems such as
copper benzoate \cite{Dender1997,Asano2000} 
or CuCl$_2$ $\cdot$ 2(dimethylsulfoxide) \cite{Kenzelmann2004}.

Earlier, a molecular orbital study based on the extended H\"uckel approach
had been performed to gain an insight in the origin of the magnetic
interaction and to study the magnetic pathway \cite{Mohri1999}.
The extended H\"uckel method 
can be viewed as a first step in the hierarchy of ab initio calculations.

In this article, we present a density functional study of this system,
in order to obtain results based on first principles calculations,
without using experimental data (apart from
the positions of the nuclei).
The target is to get an understanding of the charge- and spin distribution
by an analysis of spin density maps and a calculation of the individual
magnetic moments. By computing the energy difference between ferro-
and antiferromagnet, the exchange coupling $J$ can be extracted. 
In addition, NMR parameters such as the isotropic and anisotropic
hyperfine interaction parameters are computed and compared with
recent experimental values. The aim is to get an understanding
of the counter-intuitive experimental result, that an atom with
a relatively large distance to the magnetic ion has a larger 
isotropic shift than an atom closer to this magnetic ion.
The density functional approach allows
to obtain all these properties on equal footing.

\section{Method}
\label{methodsection}

The calculations were done with the code CRYSTAL2003 \cite{Manual03}.
This code employs a local basis set made of Gaussian type functions. For Cu,
a $[5s4p2d]$ basis set \cite{KCuF3}, 
for O a $[4s3p]$ basis set\cite{DovesiChemPhys1991} (with outermost 
$sp$ exponents of 0.5 and 0.191), 
for N as $[3s2p1d]$, for C a $[3s2p1d]$, and for H a $[2s1p]$ basis set 
was chosen; the latter three basis sets were as in \cite{Feyerherm2004}.
Full potential, all electron density functional calculations 
with the local density approximation
(LDA) and with the hybrid functional B3LYP were performed. These
calculations were done for the ferro- and antiferromagnetic state, where
the resultant solution of the Kohn-Sham
equations is an eigenstate of $S_z$, but not of $\bf S^2$. The
energy difference was therefore fitted to an Ising model, in order to
estimate the exchange coupling $J$. From the computed spin density,
the isotropic and the anisotropic hyperfine coupling parameters
are extracted.
The charge and spin of the individual atoms are obtained via the
Mulliken population analysis.

\section{Results}

\subsection{Charge and spin densities}

In table \ref{MullikenFM}, the Mulliken populations of the
ferromagnetic solution are displayed. Copper carries a charge of
$\sim$ +1.6, i.e. less than a formal charge of +2. Consequently, the total
spin is $\sim$ 0.7,
which indicates that the spin is delocalized to the neighboring atoms.
Concerning the pyrimidine ring, we notice that the nitrogen atoms
are negatively (-0.7) and the carbon atoms positively charged, so that
the ring as a whole is positively charged (0.3). The charge on
NO$_3$ is $\sim$ -0.9, and H$_2$O is approximatively neutral. 
The largest spin on the pyrimidine ring
($\sim$ 0.1) is located on the nitrogen atoms of the pyrimidine ring
which are neighbors to the copper ions. Comparing
LDA and B3LYP, we note that the LDA solution gives a slightly more
delocalized picture. This is consistent with previous findings, e.g.
\cite{MartinIllas1997,moreira,harald1,harald2}
where it was shown that LDA overemphasizes delocalization.
The spin in the pyrimidine ring is alternating up and down,
consistent with the idea of a spin polarization mechanism. 
Looking at the individual sites, for C1 and C3 essentially the
$p$-orbitals carry the spin, 
whereas for C2 the C $s$ orbital carries a little more spin.

In table \ref{MullikenAF}, the corresponding results for the
spin of the antiferromagnetic solution are displayed. The charges
are virtually identical to the charges of the ferromagnetic solution
and thus not displayed. The total spin
is similar to the ferromagnetic case for the copper atom,
and for the nitrogen atoms of the pyrimidine ring (apart from the sign,
obviously). The spin is zero for the C1 and C2 sites
due to the symmetry. 
For the C3 site, there is in the case of the LDA
a very small spin, parallel to the spin of the nearest copper, in contrast
to the ferromagnetic case, where the spin is antiparallel.

\begin{table}
\begin{center}
\caption{\label{MullikenFM}
Results from the Mulliken population analysis for the ferromagnetic
solution.}
\vspace{5mm}
\begin{tabular}{ccccc}
& \multicolumn{2}{c}{B3LYP} & \multicolumn{2}{c}{LDA}\\
atom & charge & spin & charge & spin\\ \hline
Cu   & 1.6 & 0.7 & 1.5 & 0.6 \\
\hline
N & -0.7 & 0.09 & -0.6 & 0.128\\
C1 & 0.7 & -0.01 & 0.7 & - 0.006\\
H bonded to C1 & 0.03 & 0.003 & 0.04 & 0.003\\
C2 & 0.09 & 0.01 & 0.10 & 0.02\\
H bonded to C2  & 0.02 & 0.002 & 0.03 & 0.003\\
C3 & 0.4 & -0.009 & 0.4 & -0.005\\
H bonded to C3 & 0.000 & 0.002 & -0.001 & 0.003 \\
$\Rightarrow$ Pyrimidine & 0.3 & 0.2 & 0.4 & 0.3\\
\hline
NO$_3$ & -0.9 & 0 & -0.8 & 0.01 \\ \hline
H$_2$O & -0.1 & 0.05 & -0.1 & 0.06\\
\hline
\end{tabular}
\end{center}
\end{table}

\begin{table}
\begin{center}
\caption{\label{MullikenAF}
Results from the Mulliken population analysis for the antiferromagnetic
solution.}
\vspace{5mm}
\begin{tabular}{ccccc}
& {B3LYP} & {LDA}\\
atom & spin & spin\\ \hline
Cu a,b  & $\pm$ 0.7 & $\pm$ 0.6 \\
\hline
N a,b & $\pm$ 0.08 & $\pm$ 0.10 \\
C1  &  0.000 & 0.000 \\
H bonded to C1 & 0.000 & 0.000\\
C2  & 0.000 & 0.000\\
H bonded to C2 & 0.000 & 0.000\\
C3 a,b & 0.000 & $\pm$ 0.003\\
H bonded to C3a,b & $\pm$ 0.001 & $\pm$ 0.002 \\
$\Rightarrow$ Pyrimidine & 0  & 0\\
\hline
NO$_3$ bonded to Cu a,b & 0 & $\pm$ 0.01 \\ \hline
H$_2$O attached to Cu a,b & $\pm$ 0.05 & $\pm$ 0.06 \\
\hline
\end{tabular}
\end{center}
\end{table}

\begin{figure}[ht]
\begin{center}
\includegraphics[bb= 10 -170 540 570,width=10cm,angle=0]
{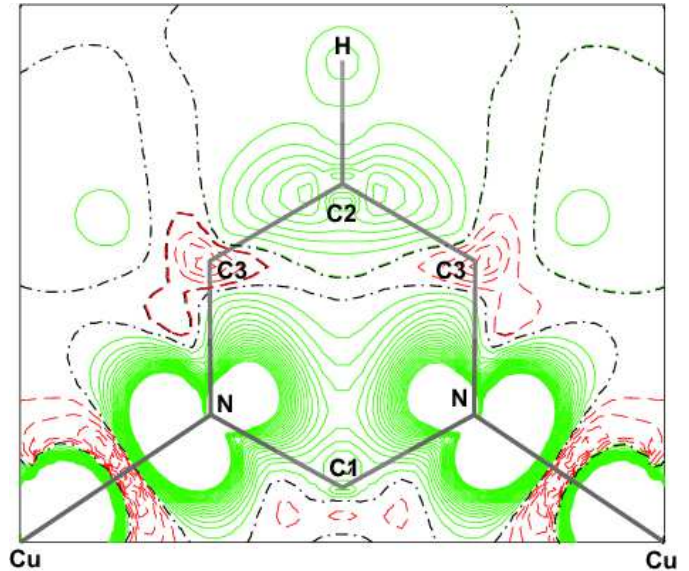}
\caption{\label{FMspindensity} Spin density of the ferromagnetic solution,
in atomic units (1/a$_0^3$, with the Bohr radius a$_0$), at the B3LYP
level. Full contour
lines represent positive spin density, dashed lines negative spin
density, and the dashed-dotted line represent zero spin density. The
lines have a distance of 0.00025/a$_0^3$.}
\end{center}
\end{figure}

\begin{figure}[ht]
\begin{center}
\includegraphics[bb= 10 -170 540 570,width=10cm,angle=0]
{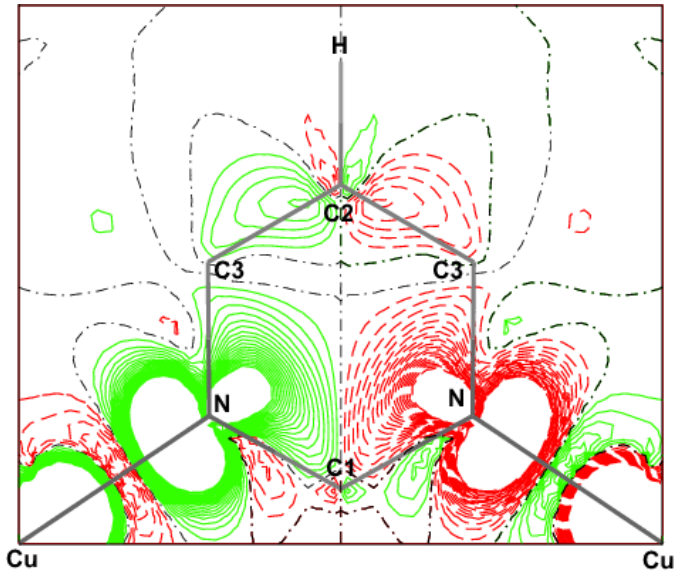}
\caption{\label{AFspindensity} Spin density of the antiferromagnetic solution,
in atomic units (1/a$_0^3$), at the B3LYP level. Full contour
lines represent positive spin density, dashed lines negative spin
density, and the dashed-dotted line represent zero spin density. The
lines have a distance of 0.00025/a$_0^3$.}
\end{center}
\end{figure}

\begin{figure}[ht]
\begin{center}
\includegraphics[bb= 10 -170 540 570,width=10cm,angle=0]
{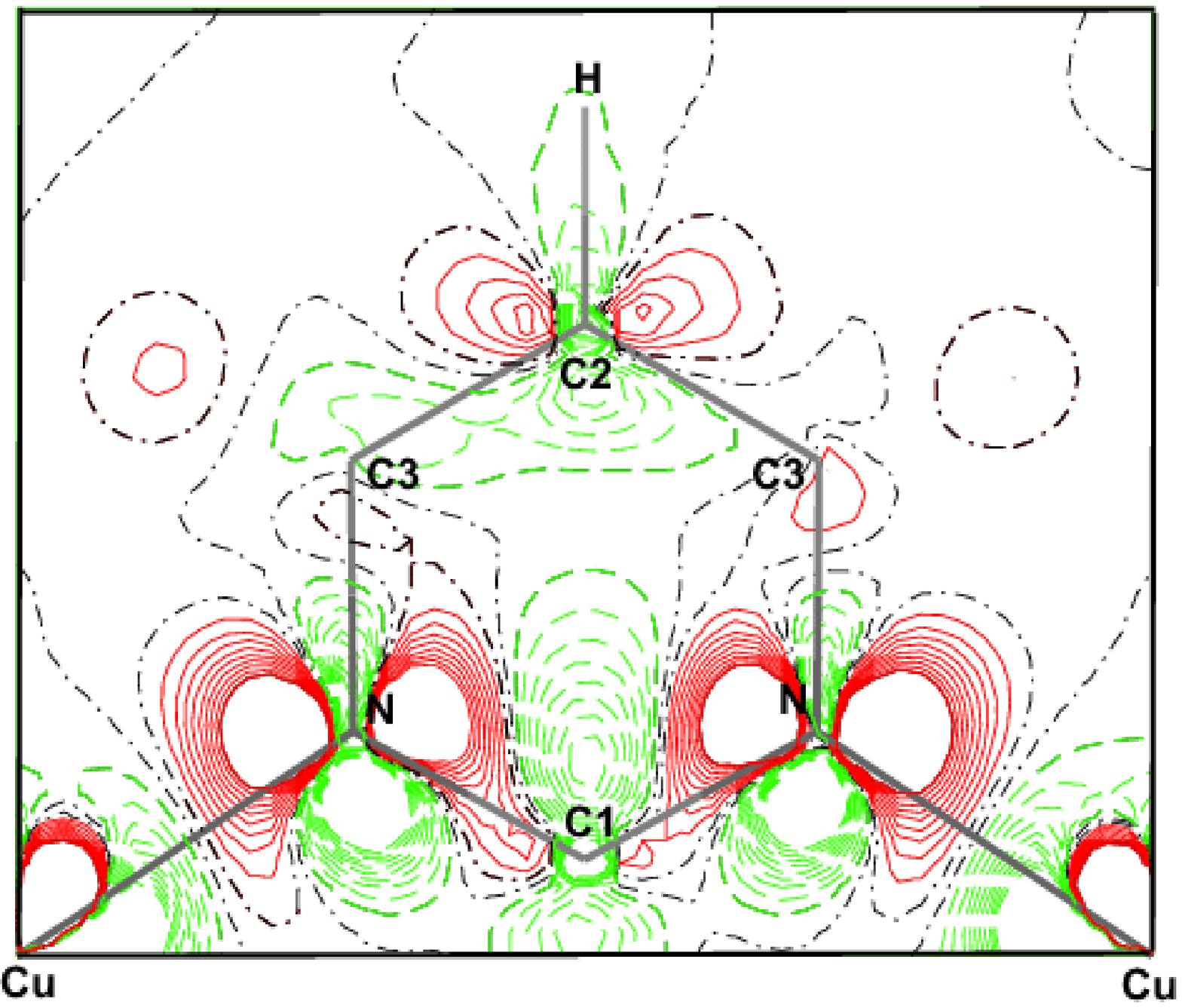}
\caption{\label{FMminusAFcharge} The difference of the charge density
of the ferromagnetic and the antiferromagnetic solution,
in atomic units ($|e|$/a$_0^3$), at the B3LYP level. Full contour
lines represent regions where the density is higher in the ferromagnetic
case, dashed lines represent regions where the density is higher in the 
antiferromagnetic
case. The dashed-dotted line corresponds to zero. The
lines have a distance of 0.00001$|e|$/a$_0^3$.}
\end{center}
\end{figure}

These results are visualized in the spin density plots for
ferromagnetic (figure \ref{FMspindensity}) and
antiferromagnetic (figure \ref{AFspindensity}) spin density, at the B3LYP
level.
It is obvious that neighboring copper and nitrogen atoms
always have parallel spin. In the ferromagnetic case,
the adjacent carbon atoms (C3) have a spin
which is antiparallel to the nitrogen atoms: antiparallel spin between
neighboring atoms reduces the Pauli repulsion.
This allows a more diffuse
charge distribution and thus reduces the energy.
Finally, the carbon atom C2
has again an antiparallel spin with respect to the neighboring carbon atoms
(C3), in order to reduce the
Pauli repulsion. In the antiferromagnetic case, the arguments hold
similarly, but additionally, the cancellation of negative and positive
spin density must be taken into account which results in a total
spin of zero or nearly zero at the carbon sites.

Comparing the ferro- and
antiferromagnetic solutions, the antiferromagnetic spin density
allows the charge densities 
of the two nitrogen atoms to stronger interpenetrate
in the center of the ring,
which is suppressed in the ferromagnetic case because of the Pauli
principle. 
This is illustrated in figure \ref{FMminusAFcharge} where it is
shown that the charge density in the ring is slightly higher for
the antiferromagnetic solution. As a whole,
this results in a stronger delocalization of the nitrogen charge
in the antiferromagnetic case and thus reduces the total energy, which
supports the antiferromagnetic coupling observed experimentally.

\subsection{Magnetic hyperfine interaction}

\begin{table}
\begin{center}
\caption{\label{Fermicontact}
Fermi contact coupling, at the level of LDA and B3LYP, versus the
experimental value, in 1/a$_0^3$, with the Bohr radius a$_0$.}
\vspace{5mm}
\begin{tabular}{ccccc}
site & LDA & B3LYP & experiment \cite{anja2005prl} \\
C1 & 0.008 &  0.003  &  0.0045\\
C2 & 0.008 &  0.007  &  0.034 \\
C3 & 0.001 & -0.0006 & -0.006 \\
\end{tabular}
\end{center}
\end{table}

From the spin densities, it is possible to compute the isotropic 
Fermi contact and the dipolar contribution to the
anisotropic hyperfine coupling. The isotropic part
is given by the spin density at the carbon nuclei, and can be
compared with the values obtained from $^{13}$C NMR \cite{anja2005prl}.
As a magnetic field is applied in the experiment, we therefore
have to use the Fermi couplings obtained with the ferromagnetic solution.
The data are displayed in Table \ref{Fermicontact}. It should
be mentioned that computing Fermi contact couplings accurately is
a notorious problem already for molecules \cite{KauppBuch}. This is even
more difficult here, as the 
spin density must be evaluated at the position of atoms which are
far away from the magnetic copper ions, i.e. transferred hyperfine
fields. The B3LYP approach
reproduces the signs of the spin densities at the different atoms 
properly: the spin density at the carbon nucleus
is positive for C1 and C2, and negative for C3. 
Note that
the total spin is negative for C1, whereas the spin density at the nucleus
is positive and even in rough quantitative agreement
with the experimental value. For the C3 site, we find a small
negative spin density at the nucleus, in qualitative
agreement with the experiment. Finally, for the C2 site, a positive
Fermi contact coupling is found which is the largest of the values
computed. Again, this is in qualitative agreement with the experiment.
Note that the value is largest at this site which
has the largest distance to the copper atoms; 
one might rather expect to find larger
Fermi couplings for atoms with shorter distances. From the spin density plots,
we find an interpretation and explanation: first, in the
pyrimidine ring, neighboring atoms have antiparallel spin, so that
we find alternating up and down spin. As the nitrogen spin is very large,
the C1 atom has a relatively small spin density because
the negative C1 spin is partially compensated by the positive nitrogen spin
which is spatially extended towards the C1 site.
In addition, the carbon spin resides essentially in the
$p$ orbital which has a node at the nucleus and thus does not contribute
to the spin density. In contrast,
at the C2 site, a spin parallel to the nitrogen spin is obtained and
mediated via both adjacent carbon atoms. The carbon $s$ orbital, 
which has a non-vanishing spin density at the
carbon nucleus, contributes slightly more to the spin for the C2 site. 
Thus, we find a relatively
large isotropic hyperfine coupling 
although this atom has the largest distance to the copper atoms
carrying the majority of the spin. 
Finally, the carbon atom at C3 has relatively small
spin with opposite sign; here again the negative spin of the carbon
is compensated by the neighboring positive spin of the nitrogen and C2.

It becomes also apparent that B3LYP fits better to the experimental
values than LDA does; the different sign for the site C3 can only be confirmed
at the B3LYP level.

\begin{figure}
\begin{center}
\includegraphics[width=8cm]{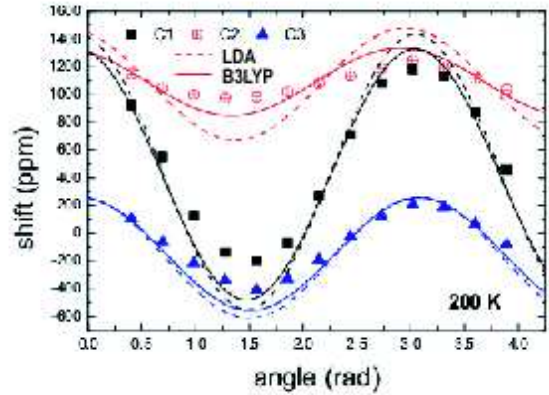}
\caption{\label{NMRanisoplot} A comparison of experimental data 
for the C1 (squares), C2 (circles) and C3 (triangles) sites
and 
a fit using experimental values for NMR chemical shift and isotropic shift,
and computed values (LDA, B3LYP) for the anisotropic shift.}
\end{center}
\end{figure}

In a next step, the components of the anisotropic hyperfine tensor are
computed. These are computed as the expectation value of the operator

\begin{eqnarray*}
T_{ijA}=\sum_{\mu,\nu}\sum_{\vec g} P_{\mu,\nu,\vec g}^{spin}\int
\varphi_{\mu}(\vec r)
\left(\frac{\vec r_A^2\delta_{ij}-3\vec r_{A_i}\vec r_{A_j}}
{r_A^5}\right) \varphi_{\nu,\vec g} (\vec r) d^3r
\end{eqnarray*}

where $\varphi_\nu$ are the Gaussian type basis functions, 
$P_{\mu,\nu,\vec g}^{spin}$
is the density matrix for the difference of up and down spin, $\vec g$
are lattice vectors, and 
$\vec r_A$ is the distance $\vec r$-$\vec A$, with $\vec A$ being the
position of the nucleus for which the anisotropic dipole
hyperfine tensor is computed.
The results of these calculations are presented in table \ref{Dipolartensor}.
A comparison is made with the results from a point lattice dipole moment
where 90 \% of the spin was allocated at the copper site, and 5 \%
at each of the two nitrogen atoms of the pyrimidine ring.
This was found to be the best fit to the measured NMR
data \cite{anja2005prl}. Such a spin distribution is
similar to the one obtained in the present work by a first principles
density functional calculation. Comparing the
dipolar tensors, we find a reasonable agreement between the
three approaches used (point dipolar model, LDA, B3LYP). This is also
demonstrated in figure \ref{NMRanisoplot}, where the computed
anisotropic hyperfine interaction tensor was used, together with
experimentally
determined NMR chemical shift and isotropic hyperfine interaction.
It can be seen that the data is reasonably well fitted, but also
the dependence on the functional becomes obvious; and B3LYP fits the
data better.

It should be mentioned that the results for the hyperfine interaction
were obtained with the basis set as described in section \ref{methodsection}.
As this property depends on the spin density of the nucleus, additional
calculations were performed with e.g. a much larger set of tighter carbon
$s$ and $p$
basis functions to describe the electron density in the region of 
the nucleus better. We found that 
the results were essentially stable with respect to 
various basis sets employed.

\subsection{Magnetic exchange coupling}

The magnetic exchange coupling is usually obtained by
computing the energy difference between two magnetic states, and fitting it to 
a model Hamiltonian. This means that the energies of at least two magnetic
states have to be computed, in order to extract $J$ from the energy
difference. 

In the case of molecules, the energy can be 
computed with accurate wave function based methods such as the
multi-reference configuration interaction scheme, multi-reference perturbation
theory and related methods
\cite{Fink1,Fink2,Graaf2001,Calzado2000,MoreiraKNiF,harald2}. 
The advantage is that the correlation
treatment is well controlled, and by taking into account more and
more determinants, a fairly accurate calculation of the exchange coupling
is possible. In addition, the wave function
can be constructed as an eigenstate of $\bf {S^2}$. On the other
hand, such calculations are very demanding in terms of memory requirement,
and are usually limited to few magnetic centers (two in most cases). 
In addition, it may
be necessary to truncate large molecules and create some embedding. 
An alternative is
to use methods such as unrestricted 
Hartree-Fock theory
or density functional theory and apply them to molecules. 

In the case of solids,
a wave-function based treatment of the periodic solid is usually
prohibitive. Thus, a possible way is to use schemes such
as Hartree-Fock theory or density functional theory which can 
be applied to solids. This approach can be performed whenever the unit cell is
not too large. 

The main downside of Hartree-Fock theory is that it
gives a too localized picture and usually strongly underestimates
the exchange coupling, often by a factor of 3: e.g. for NiO and MnO
\cite{Towler1994}, KXF$_3$ (X=Mn,Fe,Co,Ni) 
\cite{Ricart1995,Dovesi1997} or KCuF$_3$, K$_2$CuF$_4$ and Sr$_2$CuO$_2$Cl$_2$
\cite{KCuF3,Moreira2004IJQC}.

The local density approximation, on the other hand, strongly
overestimates the exchange coupling, because the charges are too
delocalized and thus exchange integrals are too large. This overestimation
is often as large as a factor of 5: e.g. for NiO 
\cite{moreira}, KCuF$_3$, K$_2$CuF$_4$ and Sr$_2$CuO$_2$Cl$_2$
\cite{Moreira2004IJQC}. Gradient corrections only slightly change this
and again, an overestimation was observed, e.g. \cite{Kortus2001,harald1}.
The situation is more difficult in systems where various couplings are
important, e.g. \cite{Park2004}.
The hybrid functional B3LYP was initially designed for molecules, but has
become very popular in solid state physics because the band gaps
obtained are in surprisingly good agreement with experiment \cite{Joe}.
It interpolates between 
Hartree-Fock theory and density functional theory,
and is now also frequently used for the calculation
of superexchange coupling constants, where it 
overestimates
exchange couplings by a factor of the order of $\lesssim$ 2, e.g. 
NiO \cite{moreira}, KCuF$_3$, K$_2$CuF$_4$ and Sr$_2$CuO$_2$Cl$_2$
\cite{Moreira2004IJQC,Feng2004}, or La$_2$CuO$_4$,
La$_2$NiO$_4$, KNiF$_3$, NiF$_2$, MnF$_2$, KMnF$_3$ \cite{Feng2004},
or FeCl$_{2}$(PM)$_{2}$ and NiCl$_{2}$(PM)$_{2}$ \cite{Kreitlow}. 
Finally, it should be mentioned that there are exceptions to these
simple rules of thumb, especially in cases where the two magnetic centers
and the bridging atom(s) show a strong deviation from an 180$^{\circ}$ angle
and approach 90$^{\circ}$,
i.e. strongly deviate from being 
on a straight line, e.g. $J_1$ in MnF$_2$ and NiF$_2$ 
\cite{Moreira2000p7816,Feng2004}, or the ferric wheel-like molecule as in
\cite{harald1,harald2}. In these cases, the coupling is small according
to the Goodenough-Kanamori rules \cite{KahnBuch} and a calculation becomes
more difficult.

A broken symmetry approach 
and subsequent spin projection was suggested as a way of obtaining eigenstates 
of $\bf {S^2}$ in the case of molecules \cite{Noodleman1981}.
Further suggestions to deal with magnetic states
were the spin-restricted open shell Kohn-Sham (ROKS) \cite{ROKS}
and the spin-restricted ensemble-referenced Kohn-Sham method (REKS)
\cite{REKS}. There is an ongoing discussion about the validity of the
various approaches, see e.g. 
\cite{Caballol1997,Illas2000,Illas2004,IllasTCA2006}.
Very recently, Ruiz et al suggested that one problem
was that the self interaction error
was taken into account twice when spin projection was applied together
with a self interaction correction \cite{Ruiz2005JCP}. This was however
challenged \cite{Adamoetal2006},
and it was argued that there was no firm theoretical
basis for this argument. 

In the case of solids, the only computationally feasible way
is to use a broken symmetry approach (without spin-projection).
The solution of the Hartree-Fock or Kohn-Sham equations
is thus in general not an eigenstate of $\bf {S^2}$, but only of $S_z$,
and the spatial symmetry is broken. 
The data can be fitted to an Ising model, and should rather
not be fitted to the Heisenberg model. This approach was actually
recommended as a 'simple yet elegant way out of this problem' 
\cite{MoreiraIllas2006PCCP}
(where 'this problem' refers to the problem described in the 
preceding paragraph).

A different way of treating solids would be to use some embedded cluster
scheme which again allows to apply the same quantum chemical methods
as in the molecular case. However, the truncation is not obvious and
poses again an approximation. 

The issue of using configuration interaction or spin-projection
was also discussed in
the context of quantum dots: systems with few electrons
are considered, and a model Hamiltonian is chosen where parameters such as the
effective mass and the dielectric constant are extracted from the
experiment. This allows to construct the wave function on the
level of configuration interaction
and as an eigenstate 
of $\bf {S^2}$, and the importance of doing so was discussed
for this class of systems, e.g.
\cite{Ellenberger,Melnikov,Yannouleas,ReimannRMP}. 
However, these systems are very
different from the one considered here: in the present work,
localized spins are considered, and
the orbital occupancy of the magnetic ions is essentially
determined by the crystal field. The ground state is thus often better
described by a single reference wave function, compared to the case
of molecules, where often a multi-reference wave function is necessary.
In the case of quantum dots, the situation is different: besides
examples where the local density approximation works surprisingly
well, e.g. \cite{Reimann2000}, there are other situations  
where a description by a single
determinant may be poor, e.g. in the case of large magnetic fields or in double
dots \cite{Melnikov}, and it becomes necessary 
to use configuration interaction schemes.

As we are interested in a uniform description of properties such as
the spin and charge density and NMR parameters, we therefore evaluated
these properties and the exchange coupling at the same level of theory.
The strength of the magnetic exchange interaction $J$ is thus computed
by fitting the energy difference between ferro- and antiferromagnet
to an Ising model: $H=\sum_i J S_{i} S_{i+1}$. The data obtained
for $J$ are a prerequisite required if one was interested in quantum
tunneling; the calculation of the anisotropy would be a further step
(which requires spin-orbit coupling).

There are two copper atoms per cell, and thus two couplings of the size $J$.
The energy difference between ferro- and antiferromagnet is thus
$\Delta E = E_{FM}-E_{AF}=2 z  S^2 J$ where $z$ is the number of
couplings per cell, i.e. 2 in this case. For $S=\frac{1}{2}$, we obtain
thus $\Delta E =J$. The computed couplings are displayed
in table \ref{DeltaEnergyJ}. At the B3LYP level, a value of 76 K is
obtained, at the LDA level a value of 603 K. The B3LYP value is 
by a factor of 2 too large, compared to the experimental value of $J$=36 K
\cite{Feyerherm2000,Yasui2001,anja2003prb}.
The LDA value is even larger, because
the LDA density is much more delocalized.
Such overestimations of computed exchange
couplings are typical for the functionals employed, as was mentioned above.
The B3LYP density is more localized and thus
a value closer to the experiment is obtained. These findings are consistent
with the Mulliken charges in tables \ref{MullikenFM} and \ref{MullikenAF}, 
where a more covalent picture was obtained with the LDA.

\clearpage
\begin{widetext}

\begin{table}
\begin{center}
\caption{\label{Dipolartensor}
Dipolar tensors at the carbon sites: the first value corresponds to 
the level of a localized dipole model, the second to LDA and the third one
to B3LYP, 
in 1/a$_0^3$. The components are given using a cartesian coordinate system.}
\vspace{5mm}

C1
\begin{eqnarray*}
\left(
\begin{array}{ccc}
0.01; 0.007; 0.006 & 0; 0; 0  & 0.022; 0.036; 0.033 \\
0; 0; 0 &  -0.011; -0.007; -0.005  & 0; 0; 0 \\
0.022;  0.036; 0.033 &  0; 0; 0  &  0.0004; -0.0006; -0.001
\end{array}
 \right)
\end{eqnarray*}
C2
\begin{eqnarray*}
\left(
\begin{array}{ccc}
0.0005; 0.0005; 0.001 & 0; 0; 0  & 0.0007; 0.01; 0.003 \\
0; 0; 0 &   0.002 ; 0.0002; -0.001  & 0; 0; 0 \\
0.0007;  0.01; 0.003 &  0; 0; 0  &  0.002; -0.0007; 0.00001 
\end{array}
 \right)
\end{eqnarray*}
C3
\begin{eqnarray*}
\left(
\begin{array}{ccc}
-0.004 ; -0.01; -0.008 & \pm 0.004; \pm 0.007; 
\pm 0.005  &  0.002; 0.01; 0.01  \\
\pm 0.004; \pm 0.007; \pm 0.005 &  0.01; 0.02; 0.02 & \pm  0.002; 
\pm 0.006; \pm 0.004 \\
0.002; 0.01; 0.01 &  \pm 0.002; \pm  0.006; \pm 0.004 &   -0.006; -0.01 ; -0.01
\end{array}
 \right)
\end{eqnarray*}
\end{center}
\end{table}

\end{widetext}

\begin{table}
\begin{center}
\caption{\label{DeltaEnergyJ}
Energy difference between ferromagnetic and antiferromagnetic solution
and exchange coupling $J$.}
\vspace{5mm}
\begin{tabular}{ccccc}
functional & $\Delta E$ ($E_h$) & $J$ ($E_h$) & $J$ (eV) & $J$ (K)\\
B3LYP & 0.00024 & 0.00024 & 0.0065 &  76 \\
LDA   & 0.0019  & 0.0019  & 0.052  & 603 \\
\end{tabular}
\end{center}
\end{table}

\section{Conclusion}

Density functional calculations on the molecule based magnet 
Cu PM(NO$_3$)$_2$  (H$_2$O)$_2$ were performed, using the
local density approximation and the hybrid functional B3LYP.
The exchange path via the pyrimidine ring was analyzed with
spin density plots and with a  Mulliken spin population. The calculations
prove a spin transfer from the Cu atom to the adjacent
nitrogen atoms as had been deduced from the NMR experiments.
The spin of the nitrogen atoms in the pyrimidine ring
is parallel to the copper spin in all
cases. In the ferromagnetic case,
the spin on the pyrimidine ring is alternating. The carbon atoms
C1 and C3 have a spin essentially in the $p$ orbitals, and
on the C2 site also a slightly larger spin is found in the carbon
$s$ orbital. In the antiferromagnetic case, the spin 
is virtually zero on all the carbon atoms. 

In the case of the antiferromagnet,
the charges of the two nitrogen atoms of the pyrimidine ring can
stronger interpenetrate. This delocalization reduces the
energy and  explains why antiferromagnetism is observed.
In addition, the isotropic and anisotropic hyperfine interaction parameters
were computed. For the isotropic parameters, a qualitative agreement
could be observed with the B3LYP functional. Especially, the
experimental result that the isotropic hyperfine coupling is largest
at the C2 site, which has the largest distance to the magnetic
ion, could be confirmed. It is suggested that this is due to the
relatively large contribution to the 
spin density from the $s$ orbital for the C2 site.
In the case of the
anisotropic dipolar hyperfine tensor, a good agreement with the experimental
data was found for both functionals, where again B3LYP performed better.
Finally, the exchange coupling was computed via the energy difference
between ferro- and antiferromagnetic state. A reasonable agreement
was found at the B3LYP level, whereas the local density approximation
results in by far too large values of $J$ 
due to an enhanced delocalization, 
which is a well known problem of exchange couplings computed 
with the LDA.

\section{Acknowledgement}
The authors would like to thank Prof. S. S\"ullow (Braunschweig) for
helpful discussions. This work has been partially supported by
the DFG SPP 1137 and contract no. KL 1086/6-1.

\end{document}